# Assessing the Robustness and Resilience of U.S. Strategic Highways: A Network Science Perspective

**Authors**:

Sukhwan Chung

Risk and Decision Science Team, US Army Engineer Research and Development Center – Environmental Laboratory, US Army Corps of Engineers
696 Virginia Rd, Concord, MA, 01742-2718, USA

Email: sukhwan.chung@usace.army.mil

ORCiD: 0009-0004-3873-8948

Daniel Sardak

Risk and Decision Science Team, US Army Engineer Research and Development Center – Environmental Laboratory, US Army Corps of Engineers
696 Virginia Rd, Concord, MA, 01742-2718, USA

Email: dansardak@gmail.com

Jeffrey Cegan

Risk and Decision Science Team, US Army Engineer Research and Development Center – Environmental Laboratory, US Army Corps of Engineers
696 Virginia Rd, Concord, MA, 01742-2718, USA

Email: jeffrey.c.cegan@usace.army.mil

Igor Linkov

Risk and Decision Science Team, US Army Engineer Research and Development Center – Environmental Laboratory, US Army Corps of Engineers
696 Virginia Rd, Concord, MA, 01742-2718, USA

Email: igor.linkov@usace.army.mil

# Abstract

Network science is a powerful tool for analyzing transportation networks, offering insights into their structures and enabling the quantification of *resilience* and *robustness*. Understanding the underlying structures of transportation networks is crucial for effective infrastructure planning and maintenance. In military contexts, network science is valuable for analyzing logistics networks, critical for the movement and supply of troops and equipment. The U.S. Army's logistical success, particularly in the "fort-to-port" phase, relies heavily on the Strategic Highway Network (STRAHNET) in the U.S., which is a system of public highways crucial for military deployments. However, the shared nature of these networks with civilian users introduces unique challenges, including vulnerabilities to cyberattacks and physical sabotage, which is highlighted by the concept of contested logistics. This paper proposes a method using network science and geographic information systems (GIS) to assess the *robustness* and *resilience* of transportation networks, specifically applied to military logistics. Our findings indicate that while the STRAHNET is *robust* against targeted disruptions, it is more *resilient* to random disruptions.

# 1. Introduction

Network science is a powerful tool for analyzing transportation networks, offering insights into their underlying structures and enabling the quantification of their resilience and robustness [1-5]. Numerous studies have focused on identifying different structures within transportation networks and understanding their characteristics and vulnerabilities [6-11]. For instance, network science has revealed that road networks and airline networks exhibit fundamentally different structures. Road networks are often planar [9, 12] and have a hierarchical mesh structure, consistently connected throughout, which provides high resilience against random disconnections [13, 14]. In contrast, air networks typically follow a scale-free structure with hubs and spokes, making them more resistant to random failures but significantly more vulnerable to targeted disruptions [13, 15]. Understanding these structural differences is crucial for effective planning, maintenance, and upgrading of transportation infrastructure.

Although the robustness and resilience of critical infrastructure have been the focus of extensive research, the definitions used to quantify these properties vary among researchers [8]. Generally, robustness refers to a system's ability to withstand disruptions without significant degradation of function [16], whereas resilience emphasizes the ability to absorb initial shocks and recover rapidly from disruptions that are often unanticipated [17, 18].

Two of the most common disruption models used when analyzing resilience and robustness are targeted and random disruptions. Targeted disruptions model deliberately planned attacks designed to maximize damage to the network, while random disruptions simulate failures occurring by chance [8]. More realistic disruption models, such as the flooding of New York City [19] or Norman, Oklahoma [20], require specific data about the disruption and their results may lack generalizability due to the specificity of the case study. As a result, studies such as [18], [21, 22] have used centrality-based targeted disruptions and random disruptions to analyze the robustness and resilience of networks, drawing generalizable conclusions.

Network science has particularly valuable applications in military contexts, where it is used for tasks ranging from cross-country transportation planning to the analysis of military communication networks [23]. In military logistics, the ability to maintain and reroute troop and supply movements under disruptions is critical. The U.S. Army's success since World War II has often been attributed to its logistical capabilities. In the critical "fort-to-port" phase, where force projection begins in the U.S., highways and railways play an indispensable role as troops, equipment, and supplies are transported from military bases to ports [24]. Studies applying network science to multi-modal logistics [25, 26] can help improve infrastructure by identifying network weaknesses.

The importance of highways in military strategy is underscored by historical precedents. The Interstate Highway System, initiated under President Dwight D. Eisenhower in the 1950s, was designed with a dual purpose: to improve civilian transportation and to serve as a critical component of national defense [27]. Eisenhower's vision was shaped by his experiences during World War II, where the efficient movement of military resources was paramount. The Strategic Highway Network (STRAHNET), defined by the Department of Defense (DoD) and the Federal Highway Administration (FHWA), comprises 62,657 miles of highways connecting military installations and ports across the U.S. [28, 29].

Despite the strategic intent behind STRAHNET's creation, the majority of highways in the United States are non-military assets. Unlike other critical infrastructures, such as military communication lines, where the military maintains exclusive control, transportation infrastructure—such as highways and railways—is shared between military and civilian users. As noted, "The Army relies on various interdependent infrastructures, the majority of which it does not own or operate" [24]. This shared nature introduces unique challenges.

Unlike purely military assets, designed and maintained with security as a primary concern, public transportation infrastructure must balance civilian needs and economic considerations. This makes such networks vulnerable to evolving threats, including cyberattacks and physical sabotage, requiring adaptive and proactive measures to ensure robustness and resilience. The concept of contested logistics highlights the potential for adversaries to disrupt these networks, causing significant operational challenges. "Peer threats possess the capability and capacity to observe, disrupt, delay, and attack U.S. forces at any stage of force projection, including while still positioned at home stations in the United States and overseas" [24].

In this paper, we propose a method that integrates network science and geographic information systems (GIS) to measure the robustness and resilience of transportation networks. This approach is applied to military logistics networks to identify vulnerabilities in the existing strategic transportation network under different types of disruptions. Our findings reveal that robustness and resilience are both desirable characteristics of transportation networks but are not necessarily interdependent. For example, the STRAHNET is more robust against targeted disruptions but more resilient to random disruptions.

## 2. Methods

In this section, we describe the sources and details of the data used, steps needed to build logistics networks using strategic highway network, and how we analyzed robustness and resilience of such network. The data used to model military logistics includes the transportation network data (STRAHNET), the locations of military installations (forts), and the locations of strategic seaports (ports). Fig. 1 summarizes the required data and steps needed to model logistics network.

*Robustness* and *resilience* in the context of transportation networks are related to the ability of the system to maintain its operational functionality and performance in the face of disruptions. A robust transportation network can withstand shocks and continue to operate unaffected, whereas resilient transportation network can effectively absorb shocks and continue to operate by rerouting traffic. In Subsubsection 2.4.4 we describe how robustness and resilience of a network can be determined.

All network analyses were performed using python (ver. 3.9), mainly using packages geopandas (ver. 0.14.3), networkx (ver. 3.2.1), and numpy (ver. 1.26.4). Visual representation of the geographic data was performed using QGIS (ver. 3.36) and contextily (ver. 1.6.2).

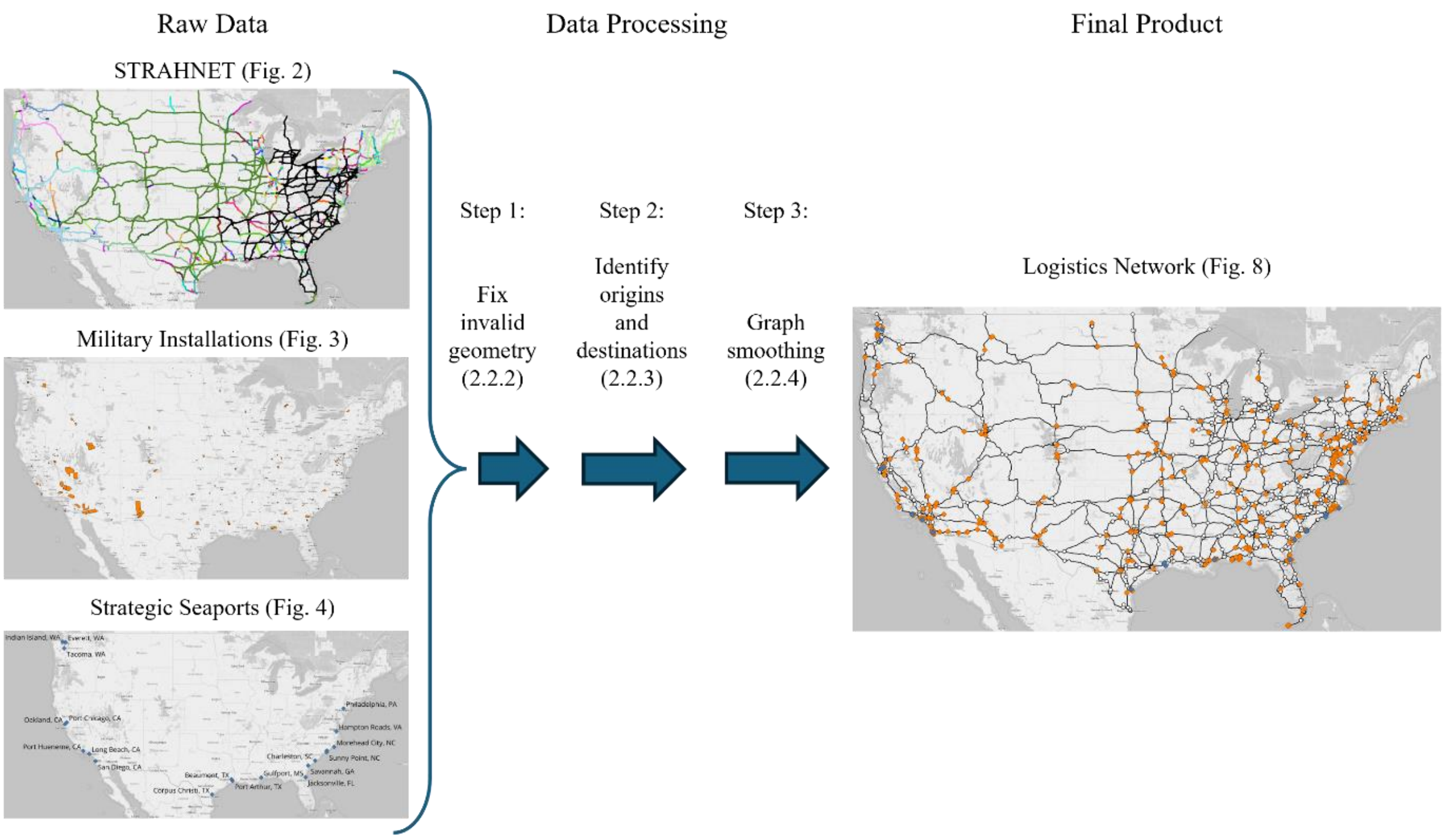

Fig. 1: Summary of data used and data processing steps to model logistics network with relevant sections.

### 2.1 Data

#### *2.1.1 Strategic Highway Network (STRAHNET)*

The raw STRAHNET data [25], [26] is a collection of interstates, non-interstate highways, and connector routes that covers not only the conterminous US (CONUS), but also includes highways in Alaska, Hawaii, and Puerto Rico. By focusing only on CONUS, the raw shape file contains 92,798 linestrings including their *id*, *geographical_shape, route_name*, *speed_limit*, *road_length*, and other information.

A continuous strip of road on a map may be represented by multiple geographical linestrings of road segments. This segmentation occurs naturally when road characteristics (such as speed limits and road names) change, but it can also occur due to data collection errors. By analyzing the STRAHNET data, a single road segment lengths ranged from 0.001 miles to 63 miles. We found that the length of a road segment calculated from the geometric shape of a linestring and the value of *road_length* attribute differs due to the projection. As a result, we used the *road_length* attribute value as the true length of a road segment. Also, whenever *speed_limit* value of a road segment was missing, we used a default value of 55 mph. An additional attribute of *travel_time* was used, where *travel_time* of an edge is defined as the edge *length* divided by the *speed limit* of the edge. Refer to Fig. 2 for a map of STRAHNET.

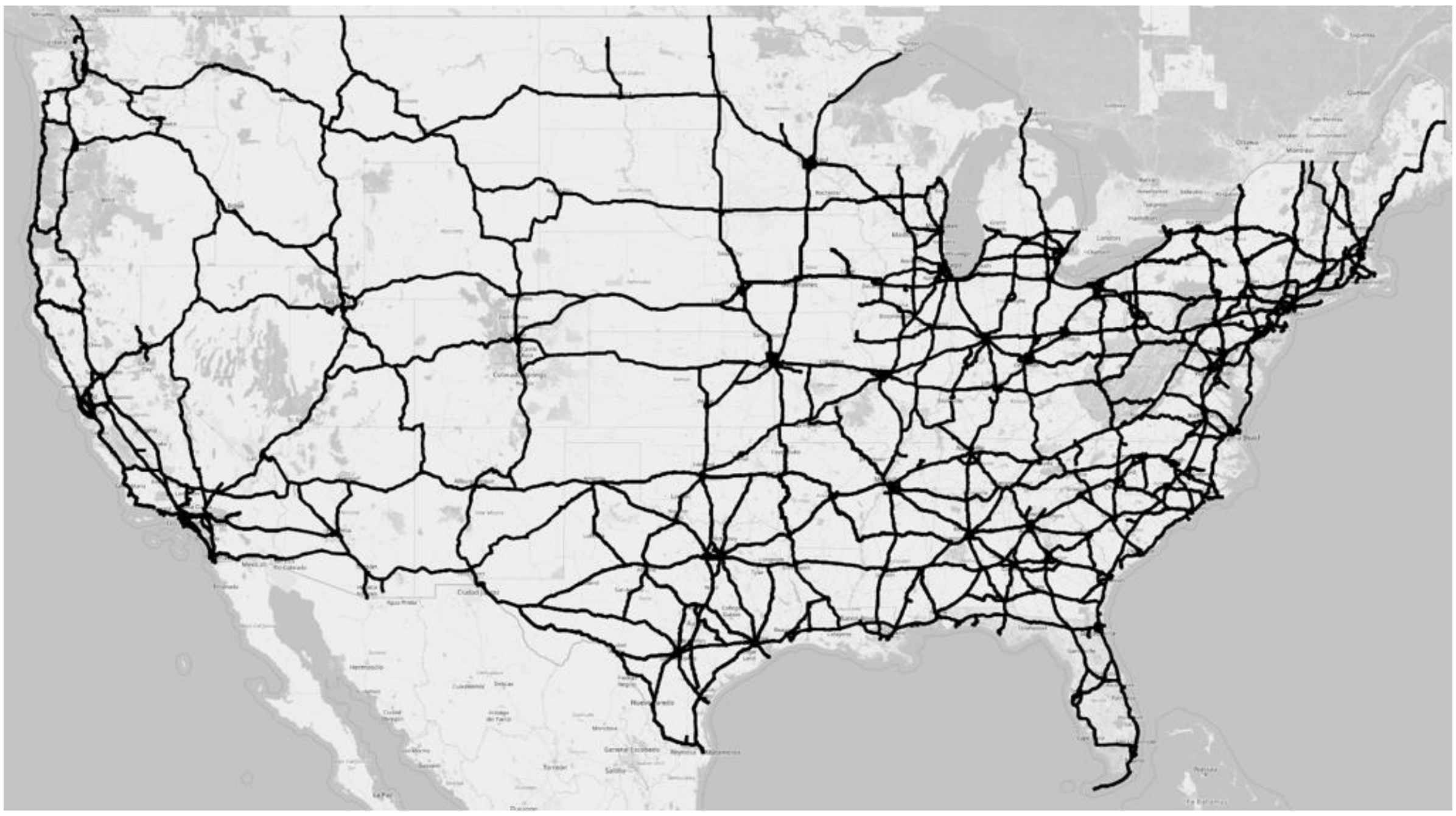

Fig. 2: Strategic Highway Network (STRAHNET)

### *2.1.2 US Military Installations*

The raw military installations data depicts the locations of the Department of Defense (DoD) installations (namely "forts"), but the boundaries do not necessarily represent the legal or surveyed land parcel boundaries [27]. It contains 765 DoD sites including national guard bases and training sites, across CONUS, Alaska, Hawaii, Puerto Rico, and Guam. By focusing only on CONUS, the raw shape file contains 673 DoD sites including their *id*, *geographical_shape, site_name*, which *department* it is under (Army, National Guard, Air Force, Navy, Marine Corps, etc.), and other information. Refer to Fig. 3 for the geographic distribution of these military installations.

### *2.1.3 Strategic Seaports*

The raw strategic seaports data depicts the locations of Strategic Commercial Seaports (namely "ports") compiled from the US Maritime Administration [28] needed to support force deployment during contingencies and other defense emergencies. It contains 18 Strategic Commercial Seaports, across CONUS, Alaska, and Guam. By focusing only on CONUS, the raw shape file contains 16 seaports including their *id*, *location, site_name*, and *capacity*. On top of the 16 commercial seaports, four additional strategic military seaports, identified by the US Government Accountability Office [29] were added to CONUS map. Refer to Fig. 4 for the geographic distribution of these seaports.

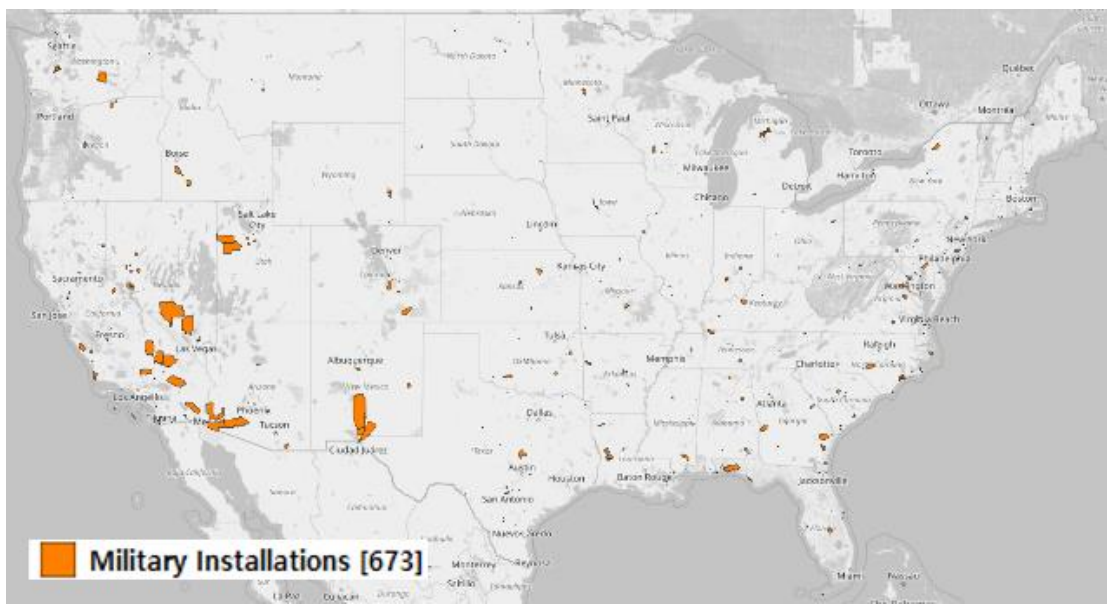

Fig. 3: Locations of military installations (“forts”) in CONUS, represented by orange polygons.

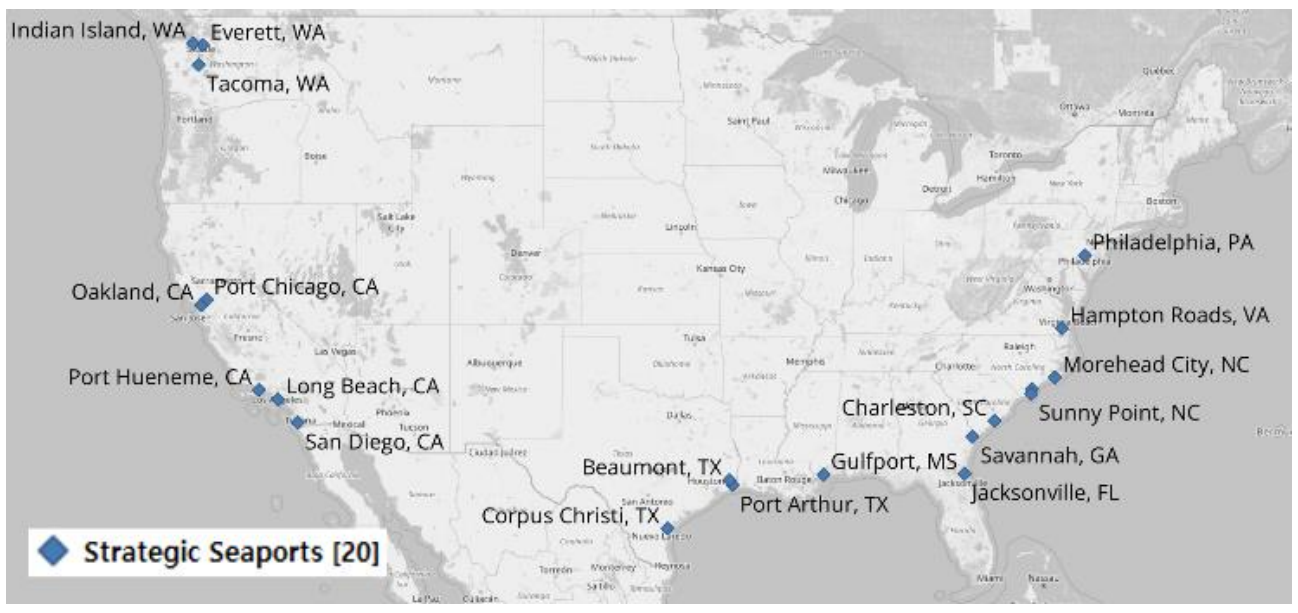

Fig. 4: Locations of strategic seaports (“ports”) in CONUS, represented by blue diamonds.

## 2.2 Logistics network representation

### *2.2.1 Transportation Graph*

The backbone of military logistics is transportation networks. To add a network science perspective of logistics on top of traditional operations research perspective and GIS perspective, a graphical representation of transportation network must be constructed. As seen in Subsubsections 2.1.1, the raw data representing strategic highways was a collection of geographical lines representing road segments. This data as it is may be useful for geographical analysis, but not suitable for network analysis. This data must be converted to a mathematical graph before it can be analyzed.

To avoid confusion, we distinguish the terms “network” and “graph” based on their subtle differences. In general, the term network and graph are interchangeably used in studies. In this paper, the term “graph”, with well-defined nodes and edges, refers to a mathematical representation of a physical “network”. In our graph representation of the highway network, the nodes represent road intersections or cul-de-sacs, and the edges represent road segments connecting the nodes. In this representation, any trip or travel must start from one node and end at another node, and movement from one road segment to another, such as turning, must happen at nodes.

The process of converting a transportation network into a transportation graph includes 3 steps.

### *2.2.2 Step 1: Fix invalid geometry*

In many cases, GIS data of roads were not generated to be interpreted as geographical graphs. Naively converting line string endpoints as nodes and linestrings as edges using data from Subsubsections 2.1.1 yields highly fragmented graphs such as Fig. 5. In reality, all the interstates and highways are connected to each other, but a naively converted highway graph is fragmented into more than 740 isolated connected components as seen in Fig. 5, where each connected component represents a group of linestrings that are connected with each other. Since the power of transportation network comes from its connectivity, this fragmentated representation of highway network cannot be used for network analysis.

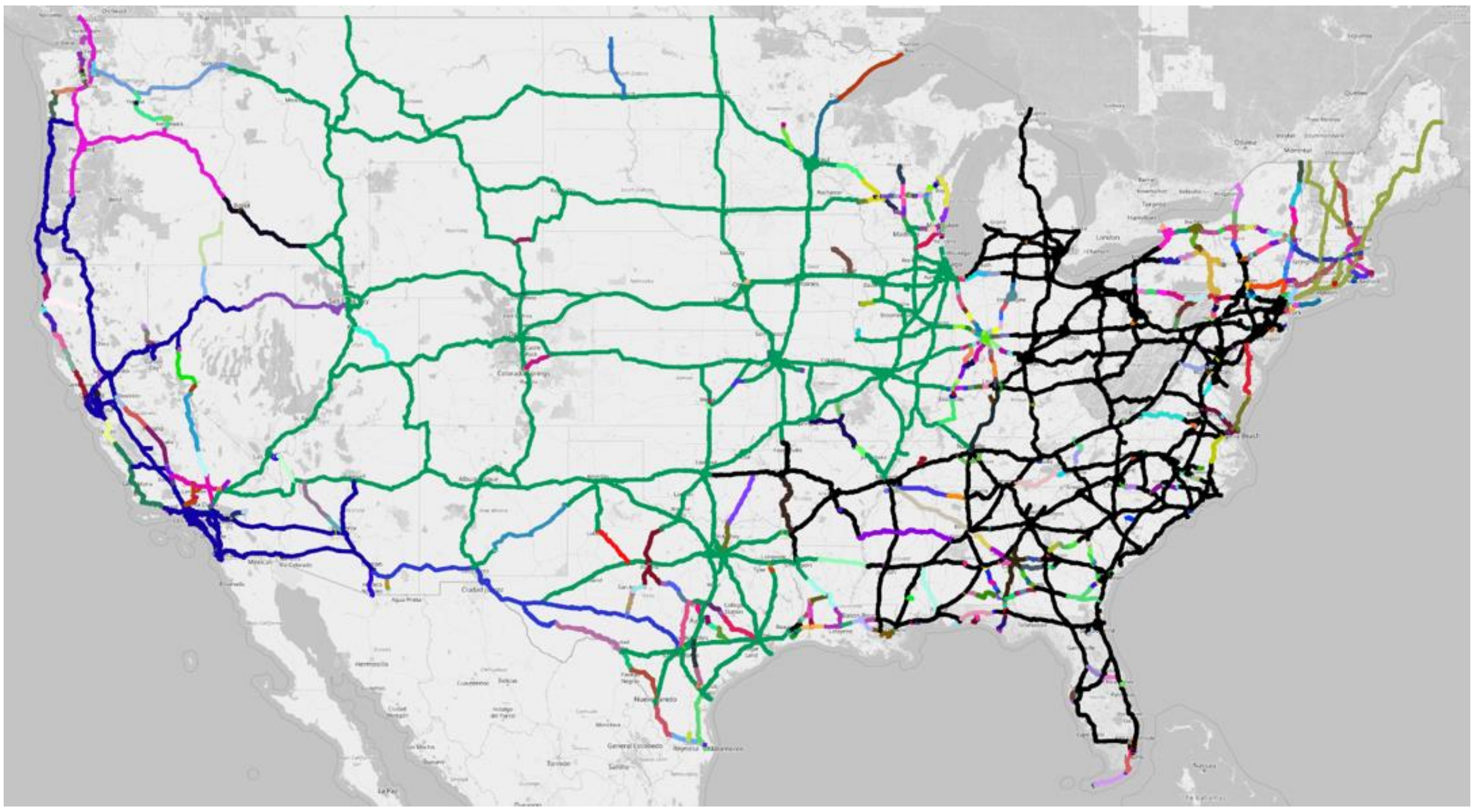

Fig. 5: A naïve conversion of the STRAHNET into a graph consisting of 92,847 linestrings. Each color represents a distinct connected component. Due to the invalid geometry issues described below, a naively converted STRAHNET graph is fragmented into more than 740 connected components.

During the graph conversion, every linestring was first converted to an edge and its end points were labelled as nodes. To connect these linestrings, if two nodes have the exact same coordinates, then the two linestrings sharing the same endpoints were connected by the shared node. During this process, fragmentation of network seen in Fig. 5 occurs because GIS representation of road segments does not perfectly align with each other, even though the physical roads they represent are connected.

These misalignments or *invalid geometries* can be categorized into three different groups: *touching*, *crossing*, and *gap*. They must be fixed before converting STRAHNET data into transportation graphs to produce a connected transportation graph. To fix these, identification of *touching points, crossing points,* and *gaps* are needed. A *touching point* is a point where one end of a linestring touches the interior of another linestring, shown in Fig. 6A. A *crossing point* is a point where the interior of one linestring crosses with the interior of another linestring, shown in Fig. 6B. Finally, a *gap* is a region where two linestrings are supposed to be connected but are not connected, shown in Fig. 6C.

The first two types of invalid geometries, *touching* and *crossing*, can be detected easily and resolved by the same method. For each linestring, identify all interior points which are touched or crossed by another linestring. Once such interior points are located, then partition the linestring at those touching and crossing points, as shown in Fig. 6D, while keeping the original geometry of the linestring. These invalid geometries can be identified using GIS analysis and can be fixed unambiguously. Here, we assumed the highway network is planar [9, 12] that the vehicles could turn at every intersection.

On the other hand, *gaps* cannot be determined just from a GIS analysis. Connection of two separated road segments cannot be determined prior to verifying with a map. Two disjoint linestring can represent parts of a continuous road strip or physically separated roads, depending on the context. Some gaps can be detected unequivocally since two line segments are separated by less than 1 $mm$ apart. On the other extreme, upon cross referencing the raw data with the STRAHNET map [26], a continuous highway segment, as long as 130 $km$, was missing from the data.

Because of this difficulty in identifying gaps, we assumed that the two road segments are connected if the separation is less than some threshold value. Specifically, detection of gaps was done in two steps: visual cross referencing and rule-based detecting. Cross-reference the STRAHNET GIS data with the official maps of the STRAHNET [26] allows some major gaps to be identified.

Once some immediately recognizable gaps were identified and connected, a rule was defined to identify smaller and less visible gaps. From all dangling nodes (nodes with degree 1), identify the nearest node that is no farther than some threshold value $\theta$. This value of $\theta$ represents the maximum distance a gap can exist. A larger $\theta$ produces an overly connected graph and a smaller $\theta$ produces overly disconnected graph than the actual transportation network.

By performing a parameter sweep on $\theta$ values ranging from $50\ m$ to $1000\ m$, a value of $400\ m$ was chosen for $\theta$. We want $\theta$ to be as small as possible without creating too many distinct connected components. Any value less than $\theta = 400\ m$ missed too many true gaps causing the graph to be fragmented, whereas any value larger than $\theta = 400\ m$ started to connect nodes that are not connected in the real world.

However, even this rule alone was not able to capture all gaps since there were gaps that exceeded $400\ m$. To identify remaining gaps, we randomly selected origin and destination pairs and calculated their travel time. If the trip took unnecessarily long detours compared to their geodesic distance, we investigated these instances to check whether the detour was due to a gap or was inevitable. We repeated this numerous times to find missing gaps and manually connected those gaps.

Fig. 6E describes how the identified *gaps* were fixed. The final product of fixing invalid geometries resulted in a connected undirected graph representation of the strategic highway network.

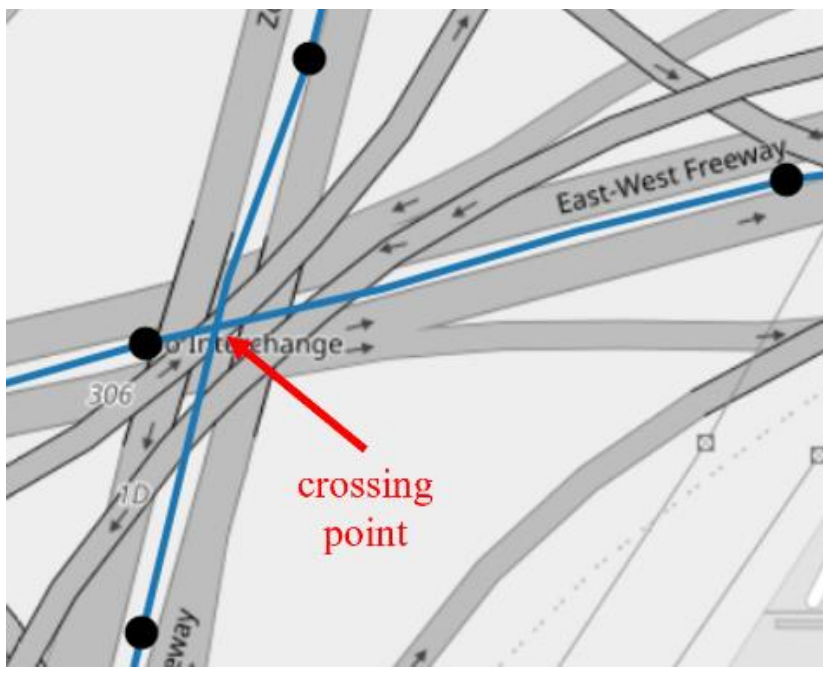

Fig. 6A: An invalid geometry *touching* occurs when one end of a linestring touches the interior of another linestring.

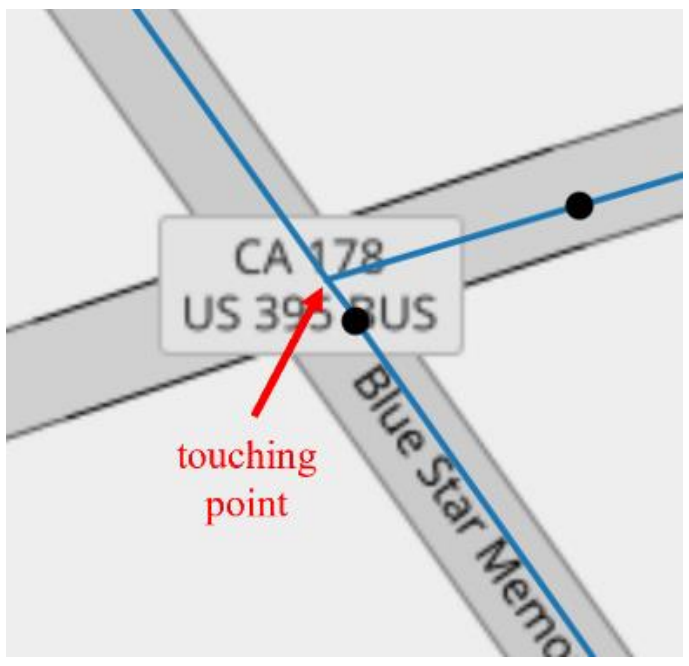

Fig. 6B: An invalid geometry *crossing* occurs when the interior of one linestring crosses the interior of another linestring.

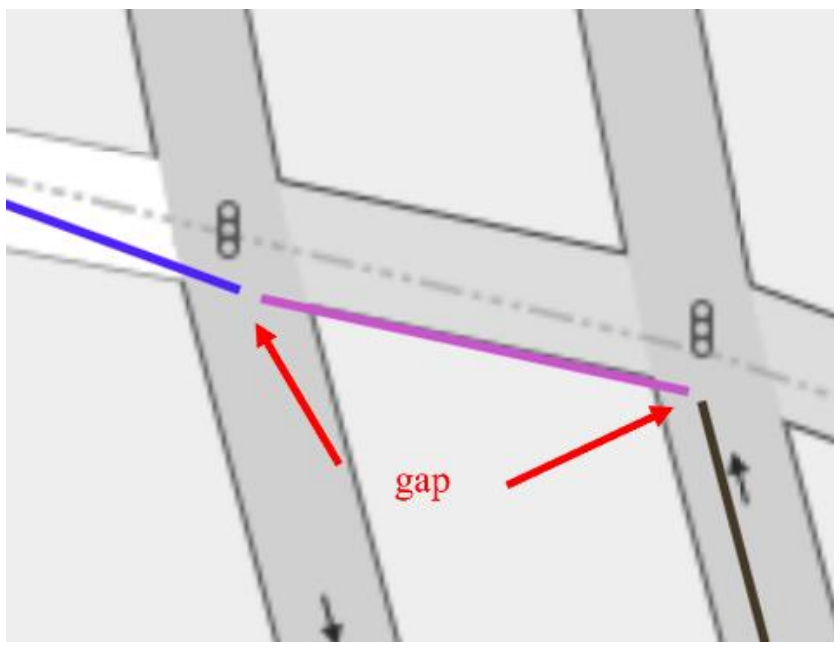

Fig. 6C: An invalid geometry *gap* occurs when the end points of road segments (part of a connected road strip) do not coincide.

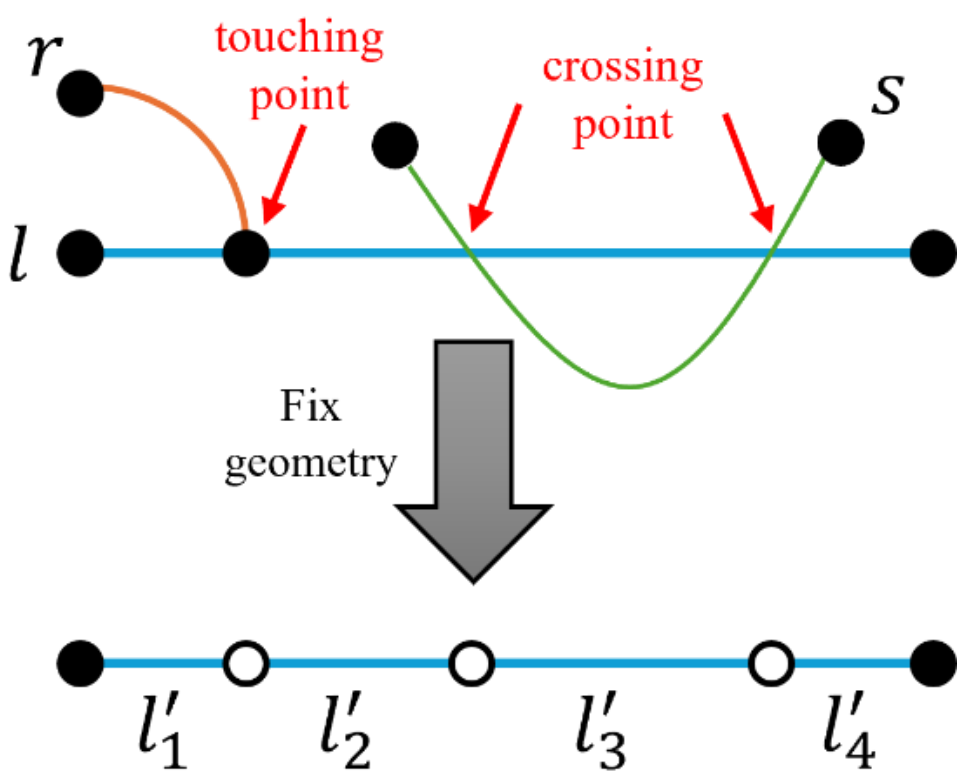

Fig. 6D: A schematic representation of how *touching* and *crossing* issues are fixed. The road segment $l$ is touched by $r$ and crossed by $s$. After detecting all the touching and crossing points, the road segment $l$ is split into contiguous pieces $l'_1$, $l'_2$, $l'_3$, and $l'_4$ such that its geographical shape and the total length $|l| = |l'_1| + |l'_2| + |l'_3| + |l'_4|$ are preserved. The black circles represent existing line ends, and the white circles represent newly added line ends.

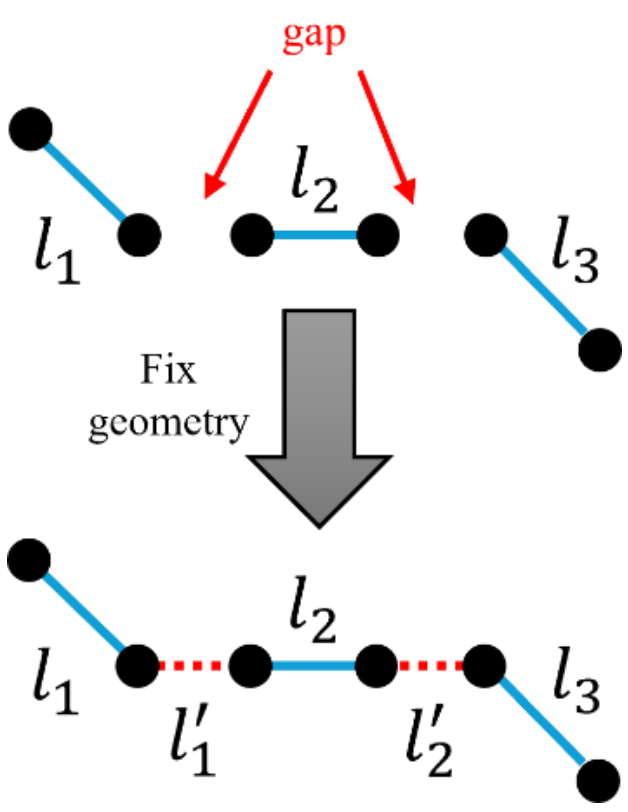

Fig. 6E: A schematic representation of how *gap* issues are fixed. The road segments $l_1$, $l_2$, and $l_3$ are part of a physically connected road. The addition of the two road segments $l'_1$ and $l'_2$ (red dashed lines) connects previously broken path.

### 2.2.3 *Step 2: Identify Origins and Destinations*

Using the military installation locations ("forts" from Subsubsection 2.1.2) as origins and the strategic seaports locations ("ports" from Subsubsection 2.1.3) as destinations, we model military logistics as finding the shortest paths from all possible forts to all possible ports. Each military installation and strategic seaport was mapped to the geographically closest point of the transportation network $G$ using Euclidean distance. If the geographically closest point on $G$ is a node, then the node is labelled as a fort node or a port node. If the geographically closest point is on an edge, then the edge is split into two at the closest point. The newly inserted point is labelled as a fort node or port node.

In some rare cases, two distinct military installations are located next to each other and are mapped to the same node in $G$. In this case, we use the same node to represent both of the installations at the same time, instead of double counting the closest nodes. Upon this identification of fort nodes and port nodes, the list of all unique fort nodes is $O$, the list of all unique port nodes is $D$, and the value $N = |O| \times |D|$ represents the number of all possible unique fort-to-port routes. Refer to Fig. 7 for a fixed STRAHNET graph with fort nodes and port nodes identified.

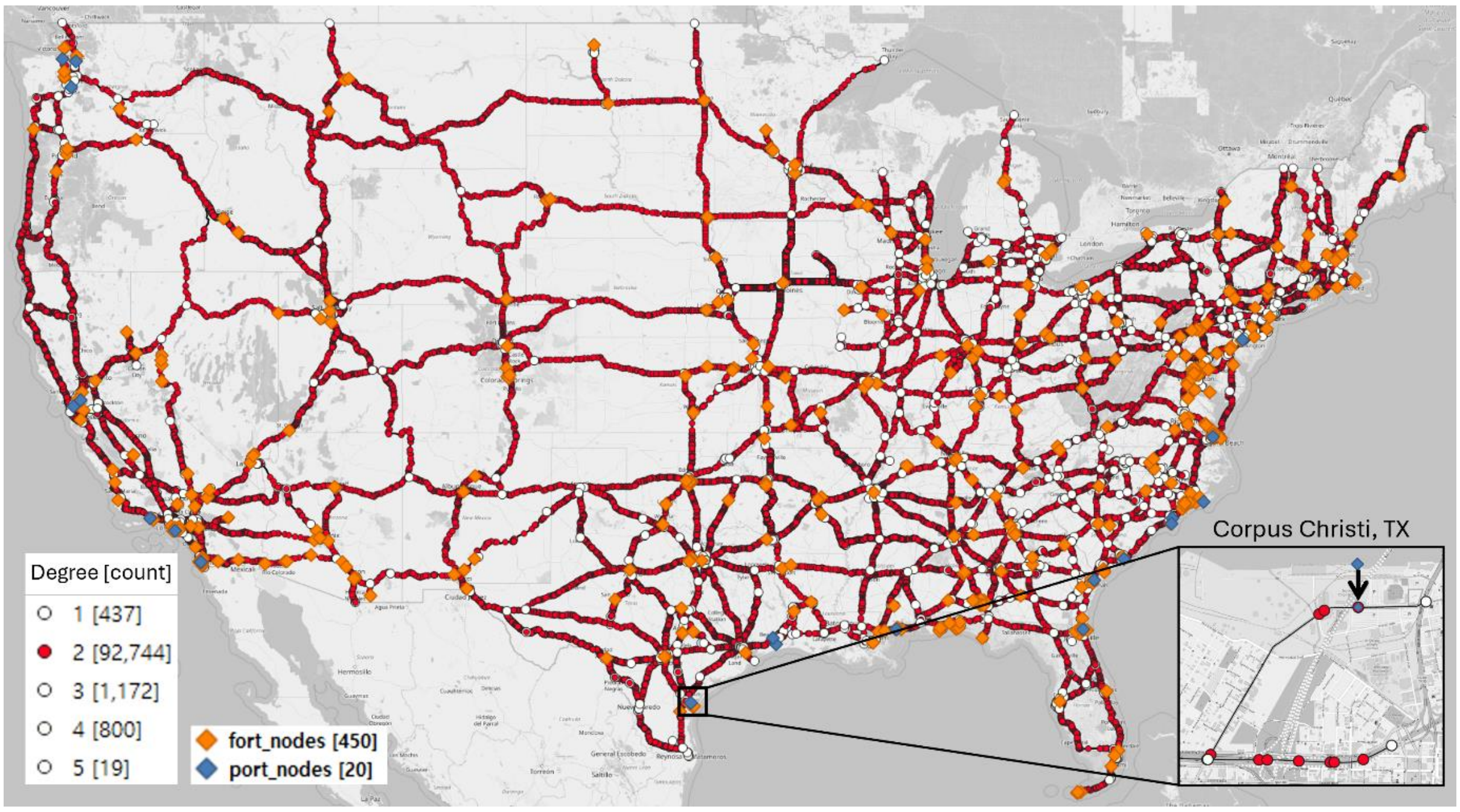

Fig. 7: Properly converted STRAHNET graph by fixing invalid geometries. Nodes with degree 2 are colored red and all other nodes are colored white. Forts and ports are represented by the closest point in the graph. Refer to the inset in black box for an example of Corpus Christi, TX.

### 2.2.4 *Step 3: Graph Smoothing*

The resulting graph after step 2 in Fig. 7 is sufficient to be used in network analysis, but it can be further simplified by *smoothing* the graph. *Graph smoothing* is the process of replacing two edges incident at the same node of degree 2 by a single new edge and removing the common degree 2 node [30]. Properly done smoothing can reduce the number of nodes and edges while maintaining the topology and attributes of the graph. To avoid removing any origin or destination node of fort-to-port routes, only the degree 2 nodes that are neither fort nodes nor port nodes were removed.

Since the edges represent the geographic shapes of roadways with different attributes, replacing two edges into one must be done carefully. Some edge attributes of the replaced edges, such as geographical shape, length, and travel time, can easily be transferred to the new edge, but other attributes such as speed limits or road names are more difficult to transfer. To simplify *smoothing*, only the *geographic shapes*, *length*, and *travel time* of the previous edges were transferred to the new edges by concatenating geographic shapes, adding the lengths, and adding travel times of the replaced edges.

Before smoothing, there were 95,172 nodes and 96,368 edges, and 97% of the nodes had degree 2 (refer to Fig. 7

lower left legend). After smoothing, there were 2,849 nodes (2.9% of the original number of nodes) and 4,045 edges (4.2% of the original number of edges) remaining, referring to Fig. 8 lower left legend. These reductions in the order (number of nodes) and size (number of edges) of the STRAHNET graph improved visual representation of the network and significantly saved computational time when performing network analyses, without altering the network topology.

The final result of STRAHNET graph using our method is shown in Fig. 8. We model logistics network as a triplet $(G, O, D)$ where $G$ is an undirected geographical graph representing transportation network, $O$ is the set of origins, and $D$ is the set of destinations.

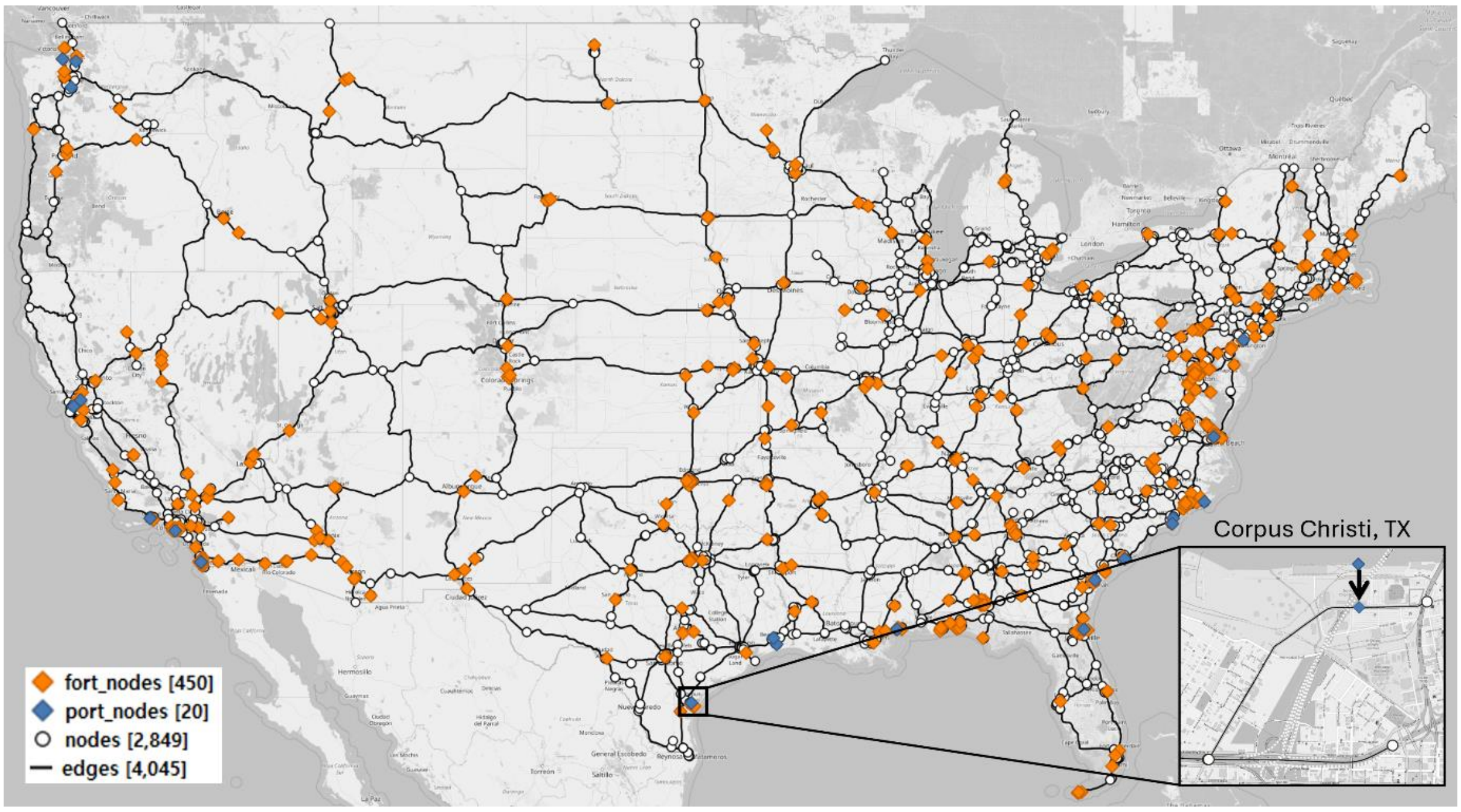

Fig. 8: The final product of converting the STRAHNET into a logistics network. Note that degree 2 nodes identified as fort nodes or port nodes were not removed by the smoothing process. Compare the black box inset in Fig. 7 and Fig. 8 of Corpus Christi, TX to see the removal of degree 2 nodes that are not fort/port nodes.

## 2.3 Network Analysis

### *2.3.1 Travel time*

In Subsubsection 2.2.2, we explained modeling military logistics is finding the fastest path for each fort-to-port route. For each origin-destination pair ($od$ pair), the shortest path is determined by the Dijkstra's algorithm minimizing the total sum of edge travel times. The travel time matrix $T$ is a $|O| \times |D|$ dimensional matrix representing travel times along the shortest paths from all possible $od$ routes. The elements of $T$, denoted $t_{od}$, represent the travel time between each origin $o \in O$ and each destination $d \in D$. Since the undisrupted transportation network $G$ is connected and $o \neq d$ for any $od$ pairs, the undisrupted travel time value $t_{od}$ is always positive and finite.

To model contested logistics, we must consider scenarios when the logistics network $G$ is disrupted where a fraction of edges was removed from $G$ (refer to Subsubsections 2.4.1 and 2.4.2 for modeling disruptions). Under disruptions, a disrupted transportation network $G'$ is less efficient than the original network $G$. Removal of an edge can never decrease the travel time for any path. As a result, the disrupted travel time matrix $T' = [t'_{od}]$ satisfies $t'_{od} \geq t_{od}$ for each $od$ pair. Furthermore, when the disruption is severe enough, the disrupted network $G'$ can be disconnected and some $od$ routes may no longer exist. In this case, we assign $t'_{od} = \infty$ to indicate disconnection of the route from $o$ to $d$.

### *2.3.2 Edge betweenness centrality*

Edge betweenness centrality or centrality for short from now on, is a measure used in network analysis to quantify the importance of an edge within a network. It represents the “number of the shortest paths that go through an edge in a graph” [31], thus capturing its role as a critical link for connectivity within the network. Mathematically, the betweenness centrality of an edge $e$ is calculated by

$$c_B(e) = \sum_{o \in O, d \in D} \frac{\sigma(o,d|e)}{\sigma(o,d)}$$

where $O$ is the set of possible origins, $D$ is the set of possible destinations, $\sigma(o,d)$ is the number of shortest paths from $o$ to $d$, and $\sigma(o,d|e)$ is the number of shortest paths from $o$ to $d$ passing through the edge $e$ [32]. The idea of shortest path centrality dates to 1977 [33] and has been widely used in network science. Usually, the betweenness centrality uses all possible node pairs ($O = D = V$) during calculation, but we modify it to use only a subset of all possible node pairs that are relevant in modeling logistics.

An edge with high betweenness centrality serves as a key channel for interactions, as many shortest paths between different node pairs pass through it. This metric is particularly useful in identifying bottlenecks in networks, where the removal or failure of a high centrality edge could significantly disrupt the overall connectivity and efficiency of the system.

We want to identify road segments that are the most important to military logistics and quantify the impacts of absence of those roads on to military logistics. Instead of calculating the centrality based on all possible node-to-node routes, we calculate the centrality based only on the fort-to-port routes. Focusing only on fort-to-port routes is computationally more efficient and more relevant to military logistics. Once the centrality measure is calculated, we can then rank edges based on their importance in fort-to-port routes. In other words, we found the most important roads for military logistics.

## 2.4 Robustness and Resilience Analysis

### *2.4.1 Disruption Modelling*

Disruptions are incidents that prevent systems from functioning properly. In this study, a disruption is represented by the removal of road segments from a transportation network, with its impact on logistics quantified by the resulting time delay on fort-to-port routes (discussed further in Subsubsection 2.4.3). In reality, many events, such as natural disasters (e.g., floods or wildfires) or manmade incidents (e.g., car crashes or terrorism), can disrupt transportation networks. When modeling such disruptions, removing nodes and/or edges are reasonable approaches; however, for simplicity, our study focuses solely on edge removals. Specifically, we model two types of disruptions—targeted and random—each with varying degrees of intensity.

The intensity of a disruption refers to the strength or magnitude of the event, while the impact relates to its broader consequences or effects on other systems of interest. In our case, disruption intensity is measured by the percentage of total road length removed, and disruption impact is quantified as the additional delays caused by the disruption (see Subsubsection 2.4.3). For our simulations, we examined disruption intensities ranging from 1% to 50% of road length, in increments of 1%.

Although it may seem unlikely for events like wildfires or acts of terrorism to simultaneously affect hundreds of miles of highways (representing only 1% of total road length), actual disruptions do not need to span the entirety of a region to cause widespread closures. For example, floods affecting a portion of a road segment can lead to the closure of miles of surrounding highways. Since the extent of highway closures due to disruptions varies significantly depending on road conditions, the severity and type of disruption, and other factors, we assume that any disruption to a road segment results in the complete removal of that segment from the network. Quantifying disruption intensity based on total road length allows for consistent comparisons of the effects of different types of disruptions on the highway system, regardless of their specific nature or origin.

### *2.4.2 Disruption Types*

Regarding the two types of disruptions, targeted disruptions involve the removal of road segments based on their importance, as measured by centrality. This approach models disruptions deliberately caused by malicious agents. During a targeted disruption, road segments with the highest centrality values are removed first, continuing until the total percentage of removed road length matches the desired disruption intensity.

In contrast, random disruptions involve the removal of road segments in a stochastic manner, such that the total percentage of removed road length equals the desired intensity. These disruptions model unpredictable events, such as vehicle collisions or tree falls. Since random disruptions are inherently stochastic, we perform 100 iterations for each intensity level to capture their statistical behaviors and variability.

#### *2.4.3 Impact Quantification*

Disruptions in transportation networks can cause negligible to serious impacts to logistics. We focused on the time delays in each path caused by disruptions. Referring to Subsubsection 2.3.1, travel times for all paths before and after a disruption were collected in matrixes $T$ and $T'$, respectively, and the elements satisfy $0 < t_{od} < \infty$ and $0 < t'_{od} \leq \infty$ such that $t_{od} \leq t'_{od}$ for each $od$ pair where $t'_{od} = \infty$ indicates disconnection.

By measuring time delays before and after the disruption, we can quantify the impact of the disruption on each path. Here, we use the relative time difference $\Delta t_{od} \coloneqq \frac{t'_{od} - t_{od}}{t_{od}}$ measured in percent increase in time with respect to the undisrupted travel time, to quantify the impact of the disruption. Using $\Delta t_{od}$, we classify the impact of the disruption on each path into three categories where the path $od$ is

1) *unaffected* if $\Delta t_{od} = 0$,
2) *delayed* if $0 < \Delta t_{od} < \infty$, and
3) *disconnected* if $\Delta t_{od} = \infty$.

This trichotomy of paths satisfy the relation $N = N_{unaffected} + N_{delayed} + N_{disconnected}$, where $N = |O| \cdot |D|$ is the total number of paths, and $N_{unaffected}$, $N_{delayed}$, and $N_{disconnected}$ represent the number of unaffected, delayed, and disconnected paths, respectively. Equivalently, we use the normalized number of paths

$$1 = \frac{N_{unaffected}}{N} + \frac{N_{delayed}}{N} + \frac{N_{disconnected}}{N} = n_{unaffected} + n_{delayed} + n_{disconnected}$$

to represent our findings (Fig. 13A and Fig. 13B).

#### *2.4.4 Robustness and Resilience*

Since the primary objective of a logistics network is to connect origins to destinations, the percentage of disconnected routes serves as a measure of system failures. Once connections between points are established, the secondary objective is to maximize efficiency, which is measured by minimizing travel times. Achieving this secondary objective relates to the percentage of delayed paths, while not ideal, still fulfill the primary objective of connectivity.

Considering these two objectives, a robust network satisfies both primary and secondary objectives under disruptions by maintaining a high proportion of unaffected paths. In contrast, a resilient network prioritizes the primary objective by preserving connectivity, even at the expense of efficiency. This is achieved by absorbing the initial shock of disruptions, converting some unaffected paths into delayed paths rather than allowing them to become disconnected.

# 3. Results

## 3.1 Travel Times

After calculating shortest paths and travel times for all possible fort-to-port routes using undisrupted transportation network, the travel times can be aggregated to show their distribution (Fig. 9). There were 9000 possible fort-to-port routes (450 forts, 20 ports) using the highways. The longest highway route identified was from Naval Air Station Key West (FL) to Naval Magazine Indian Island (WA) taking 55 hours to complete 5849 km.

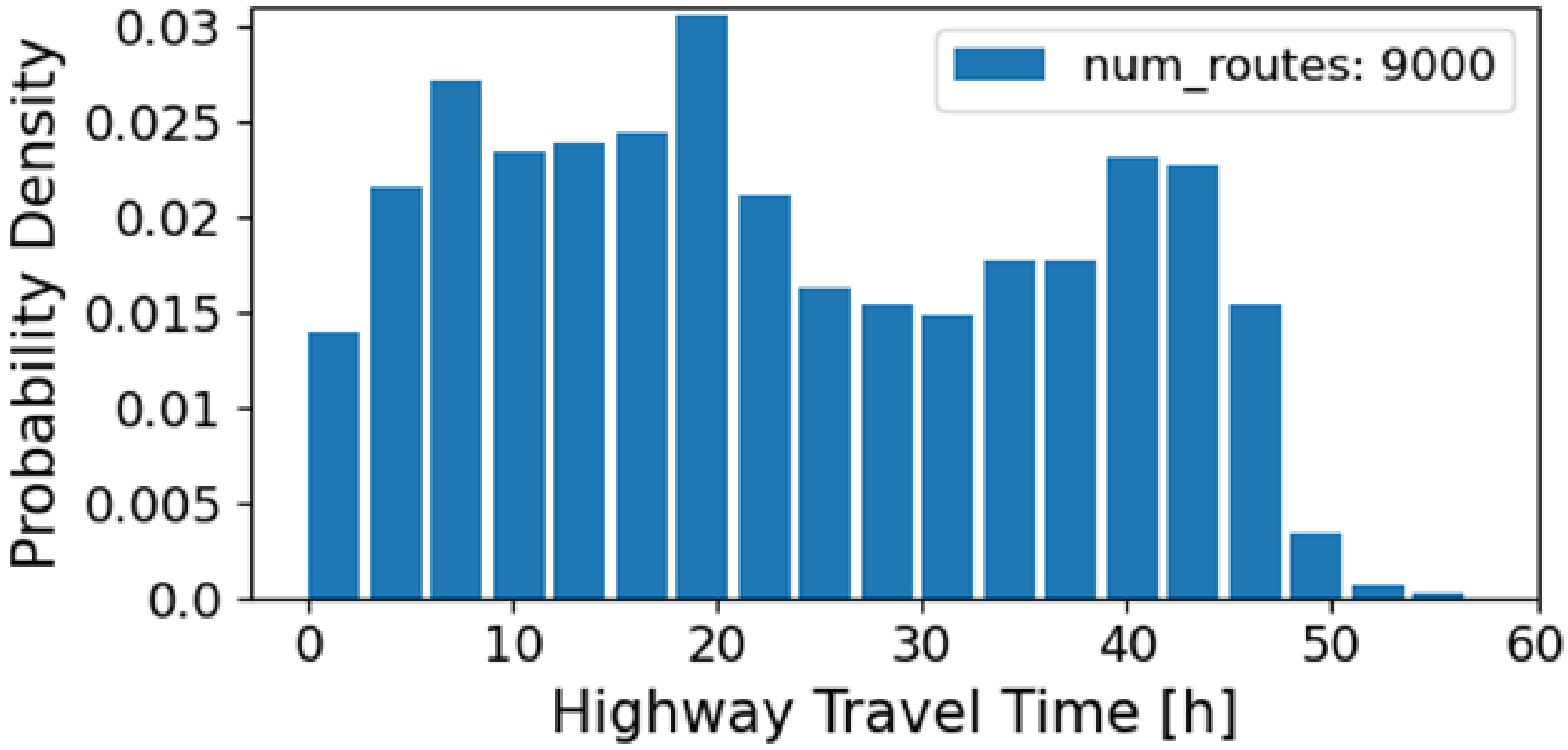

Fig. 9: Travel time distribution of fort-to-port routes using highways.

### 3.2 Important Road Segments

Once the shortest paths for all possible fort-to-port routes were calculated, centrality value for each edge was calculated (Fig. 10). Low centrality value indicates the route was rarely used for logistics and high centrality value indicates that many fort-to-port logistics rely on that particular road segment. The roads connecting Washington to New Jersey (I-90 and I-80 from west to east), and Southern California to Georgia and Carolinas (I-10 and I-20 from west to east) correspond to the highest centrality values (refer to Fig. 11A 10% disruption plot for a visually clearer representation). This tendency of high centrality values being distributed in the west-to-east direction rather than the north-to-south direction is due to the tendency of strategic seaports being distributed on the west and east coasts rather than on the north and south borders.

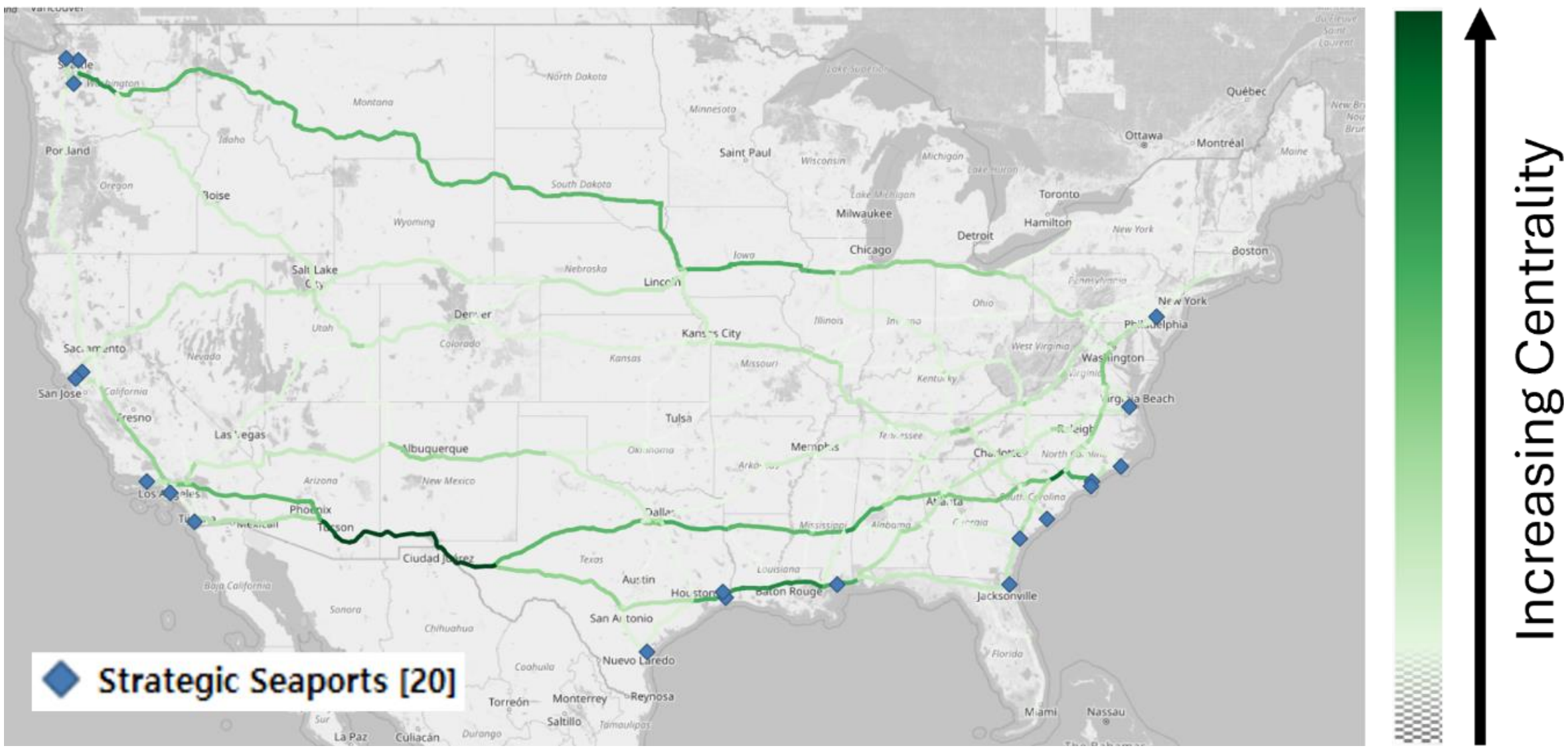

Fig. 10: Edge betweenness centrality values for the STRAHNET.
Edges with higher centrality values are highlighted by darker green and lower centrality values are transparent.

### 3.3 Disruptions

As described in Subsubsection 2.4.1 and 2.4.2, we simulated targeted and random disruptions for the STRAHNET. In Fig. 11, different scenarios in which the STRAHNET can be disrupted are displayed. Since the targeted disruptions are deterministic, stronger disruptions extend from the existing disruptions. On the other hand, random disruptions are simulated statistically independent of each other, and an ensemble of 100 disruptions are used to draw inferences for each disruption intensity.

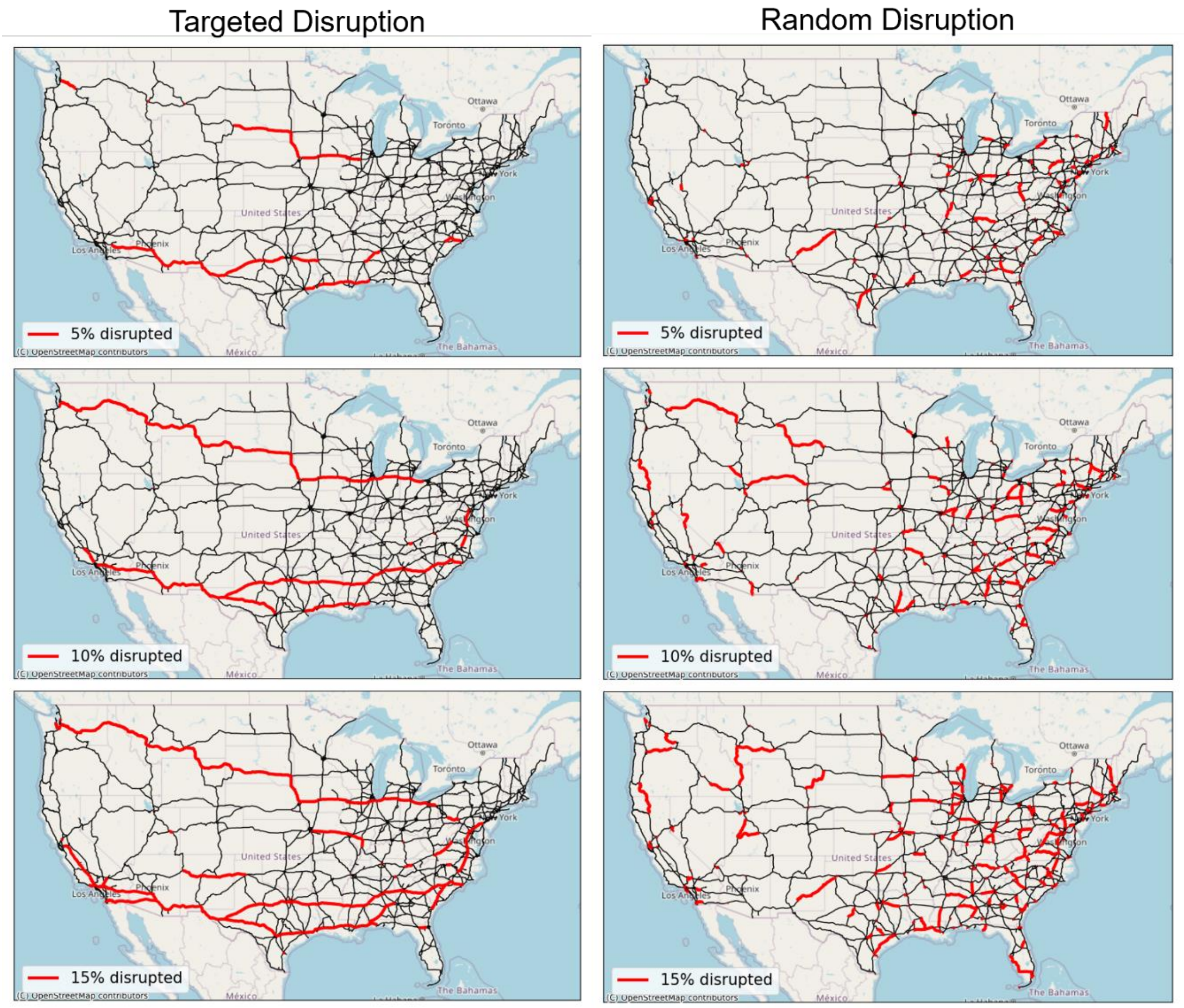

Fig. 11A: Examples of targeted disruption. Fig. 11B: Examples of random disruption.
Disrupted road segments are highlighted in red.
Highlighted edges may seem to represent more than the described intensity due to their exaggeration.

# 4. Discussions

## 4.1 Resilience

Based on the trichotomy described in Subsubsection 2.4.3, Figures 12A and 12B illustrate how the path categories change as disruption intensities increase. Under any disruption scenario, the number of unaffected paths (blue lines) decreases monotonically, and the number of disconnected paths (red lines) increases monotonically, as disruptions inherently remove edges. However, the number of delayed paths (yellow lines) can either increase or decrease depending on the relative rates at which unaffected paths disappear, and disconnected paths emerge. In both Figures 12A and 12B, the network absorbs the initial shock of disruptions by converting unaffected paths into delayed paths, preventing immediate disconnection.

Under targeted disruptions, following the initial shock, the network enters a stable phase, observed between 2% and 7% disruption intensity under targeted disruptions, where relatively minor changes occur in the path categories. Once this stable phase ends, the network experiences a secondary shock between 7% and 11%, leading to a sharp increase in disconnections starting at 13% disruption intensity.

In contrast, the network responds more gradually to random disruptions. It exhibits greater resilience to random disruptions, particularly under lower disruption intensities, which typically correspond to the initial phase of disruptions. This is because the network's capacity to delay routes reaches 75% under random disruptions, compared to 60% under targeted disruptions. While this capacity to delay does not inherently indicate better shock absorption—since some paths may remain unaffected—a comparison of disconnected paths provides further insight. At 3% targeted disruption intensity, 20% of paths are already disconnected, whereas the same level of disconnection occurs at 6% random disruption intensity. This suggests that, under random disruptions, the network absorbs initial shocks more effectively by delaying paths rather than allowing immediate disconnections. As a result, the network demonstrates greater resilience under random disruptions.

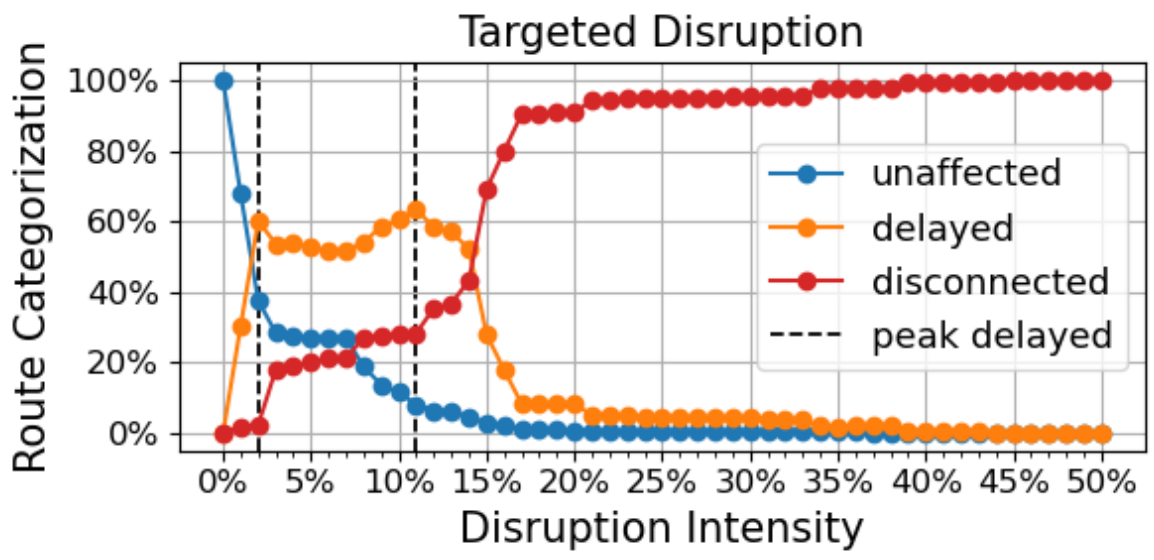

Fig. 12A: Change of $n_{unaffected}$, $n_{delayed}$, and $n_{disconnected}$ under targeted disruptions. Two dashed vertical lines indicate where $n_{delayed}$ peaked.

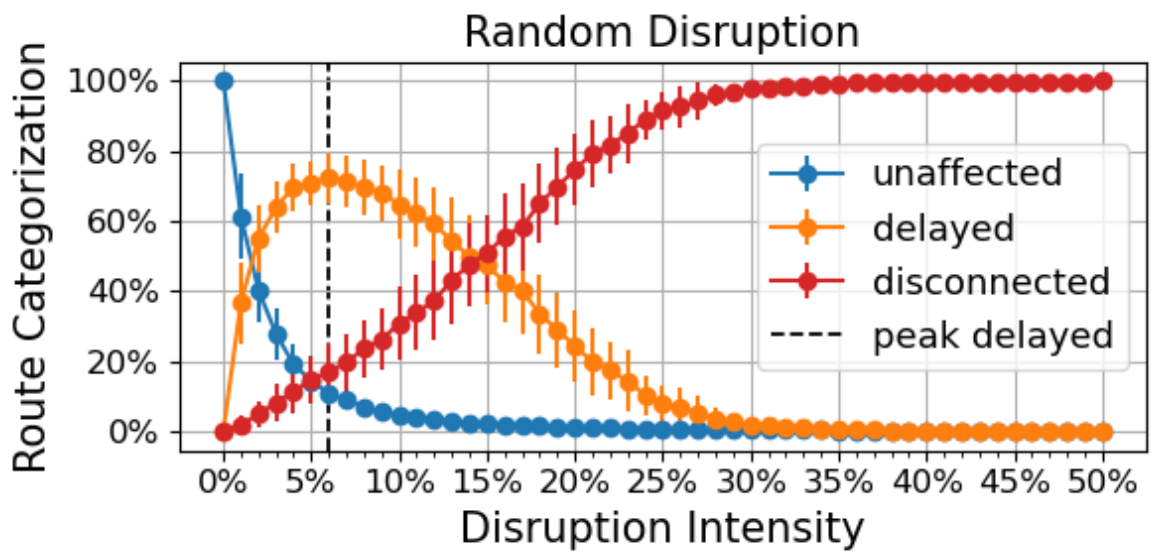

Fig. 12B: Change of $n_{unaffected}$, $n_{delayed}$, and $n_{disconnected}$ under random disruptions. The dashed vertical line indicates where $n_{delayed}$ peaked.

## 4.2 Robustness

To evaluate the robustness of the STRAHNET against disruptions, we examine how the number of unaffected routes changes, as shown in Fig. 13 and described in Subsubsection 2.4.4. In a strict sense, the STRAHNET fails to demonstrate robustness under either targeted or random disruptions, as it is unable to maintain most of its unaffected paths (approximately 70%) even under very low disruption intensities (3%).

However, unlike the monotonic decrease in unaffected paths observed under random disruptions, the STRAHNET exhibits some degree of robustness under targeted disruptions between 3% and 7% disruption intensity, maintaining a relatively stable number of unaffected paths during this range.

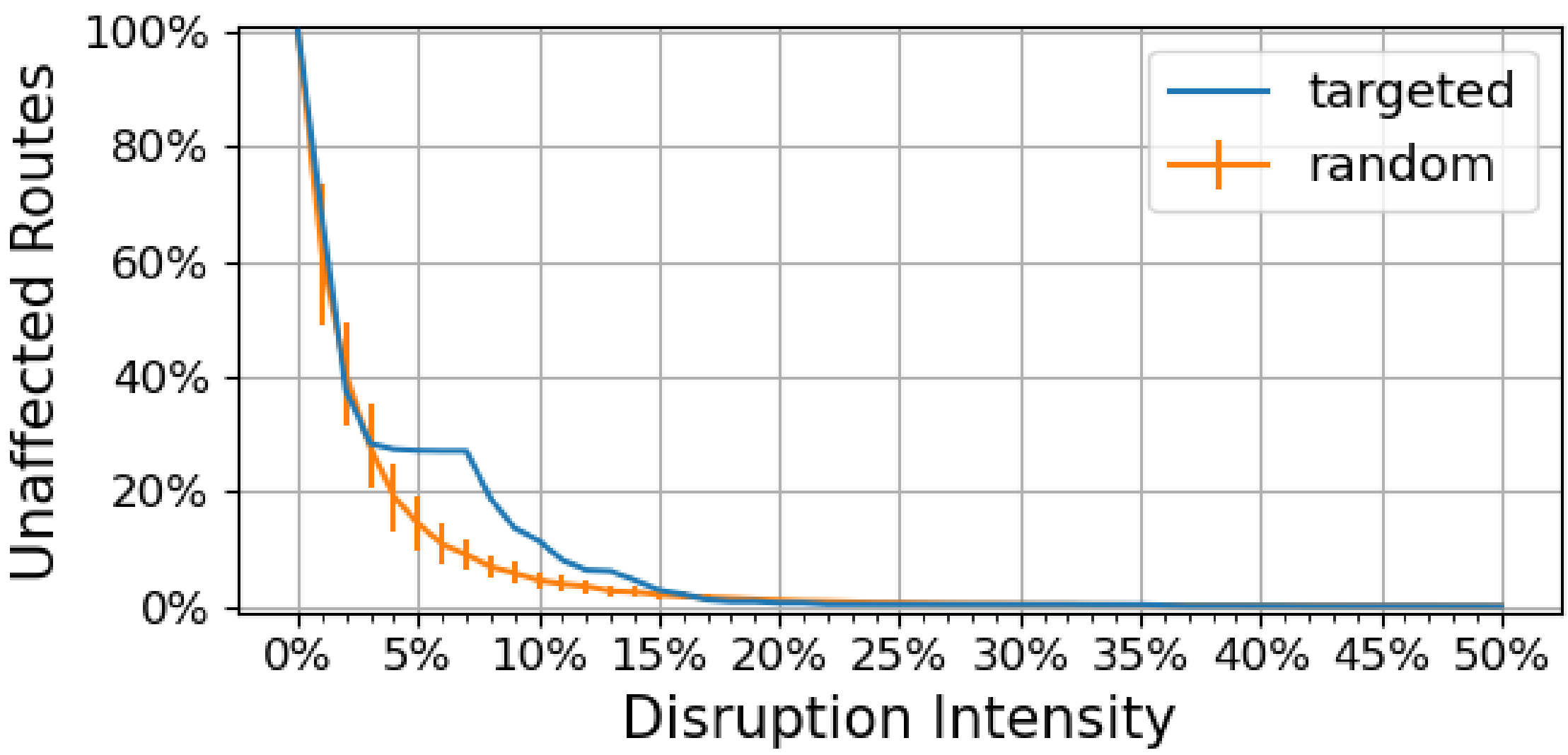

Fig. 13: Impacts of targeted disruptions (blue) and random disruptions (orange) on the number of unaffected paths on the STRAHNET. On average, random disruptions monotonically decreased the number of unaffected routes, whereas targeted disruptions has a period of no effect (3%-7% intensity).

# 5. Conclusions

In this paper, we presented a methodology that integrates network science with existing GIS perspectives to analyze the *robustness* and *resilience* of transportation networks, particularly in the context of military logistics. While numerous studies have examined the structure of various transportation networks under different types of disruptions, these studies predominantly focused on robustness through percolation analysis involving edge addition and removal. Building on existing research, we proposed a method to quantify both *resilience* and *robustness* by evaluating the difference between *delayed* and *disconnected* routes under various disruption scenarios. Applying our methodology to the STRAHNET and military logistics, we found that, although *robustness* and *resilience* are both desirable characteristics, they are not necessarily correlated. Specifically, while the STRAHNET demonstrates slightly greater *robustness* against targeted disruptions, it exhibits higher *resilience* to random disruptions.

This study has limitations, including an oversimplified logistics model. We modeled military logistics as the shortest paths between forts and ports, but in practice, military logistics is more complex, involving variables such as throughput, types of shipments, mode of transportation, cost, and capacity. Additionally, it is unlikely for a large number of troops and equipment to be transported from Florida to Washington solely by trucks. In reality, other transportation modes, such as railways, airways, and riverways, are also used for logistics, and there are many types of potential disruptions. Future work will address these limitations by considering different transportation networks, such as the STRACNET or multimodal transportation systems, as well as analyzing the impact of various disruptions, including wildfires and floods, on the robustness and resilience of transportation networks.

# Acknowledgements

We would like to thank Dr. Madison Smith, Dr. Maksim Kitsak, Dr. Andrew Jin, and Stephanie Galaitsi for their helpful feedback and comments.

**Funding:** This work was supported by the US Army Corps of Engineers, Engineer Research and Development Center FLEX program on compounding threats.

The views and opinions expressed in this article are those of the individual authors and not those of the US Army or other sponsor organizations.

Preparation of this manuscript utilized the services of ChatGPT and Microsoft Copilot Pro, which both are based on the GPT-4 architecture developed by open AI, for proofreading this document.

The authors declared no potential conflicts of interest with respect to the research, authorship, and/or publication of this article.

# Author Contributions

**The authors confirm contribution to the paper as follows: study conception and design: Sukhwan Chung, Igor Linkov; data collection: Sukhwan Chung; analysis and interpretation of results: Sukhwan Chung, Dan Sardak, Igor Linkov; draft manuscript preparation: Sukhwan Chung, Dan Sardak, Jeffrey Cegan.**

**All authors reviewed the results and approved the final version of the manuscript.**

## References

[1] D. Li, Q. Zhang, E. Zio, S. Havlin, and R. Kang, “Network reliability analysis based on percolation theory,” *Reliability Engineering & System Safety*, vol. 142, pp. 556–562, Oct. 2015, doi: 10.1016/j.ress.2015.05.021.

[2] M. Li, R.-R. Liu, L. Lü, M.-B. Hu, S. Xu, and Y.-C. Zhang, “Percolation on complex networks: Theory and application,” *Physics Reports*, vol. 907, pp. 1–68, Apr. 2021, doi: 10.1016/j.physrep.2020.12.003.

[3] S. Dong, H. Wang, A. Mostafizi, and X. Song, “A network-of-networks percolation analysis of cascading failures in spatially co-located road-sewer infrastructure networks,” *Physica A: Statistical Mechanics and its Applications*, vol. 538, p. 122971, Jan. 2020, doi: 10.1016/j.physa.2019.122971.

[4] Y. Deng, S. Liu, and D. Zhou, “Dependency cluster analysis of urban road network based on percolation,” *Transportation Research Part C: Emerging Technologies*, vol. 154, p. 104264, Sep. 2023, doi: 10.1016/j.trc.2023.104264.

[5] B. Abdulla, A. Kiaghadi, H. S. Rifai, and B. Birgisson, “Characterization of vulnerability of road networks to fluvial flooding using SIS network diffusion model,” *J Infrastruct Preserv Resil*, vol. 1, no. 1, p. 6, Dec. 2020, doi: 10.1186/s43065-020-00004-z.

[6] S. Reza, M. C. Ferreira, J. J. M. Machado, and J. M. R. S. Tavares, “Road networks structure analysis: A preliminary network science-based approach,” *Ann Math Artif Intell*, vol. 92, no. 1, pp. 215–234, Jan. 2024, doi: 10.1007/s10472-022-09818-x.

[7] Y. Duan and F. Lu, “Structural robustness of city road networks based on community,” *Computers, Environment and Urban Systems*, vol. 41, pp. 75–87, Sep. 2013, doi: 10.1016/j.compenvurbsys.2013.03.002.

[8] P. Y. R. Sohouenou, P. Christidis, A. Christodoulou, L. A. C. Neves, and D. L. Presti, “Using a random road graph model to understand road networks robustness to link failures,” *International Journal of Critical Infrastructure Protection*, vol. 29, p. 100353, Jun. 2020, doi: 10.1016/j.ijcip.2020.100353.

[9] F. Xie and D. Levinson, “Measuring the Structure of Road Networks,” *Geographical Analysis*, vol. 39, no. 3, pp. 336–356, Jul. 2007, doi: 10.1111/j.1538-4632.2007.00707.x.

[10] A. Haznagy, I. Fi, A. London, and T. Nemeth, “Complex network analysis of public transportation networks: A comprehensive study,” in *2015 International Conference on Models and Technologies for Intelligent Transportation Systems (MT-ITS)*, Budapest, Hungary: IEEE, Jun. 2015, pp. 371–378. doi: 10.1109/MTITS.2015.7223282.

[11] D. Das, A. Kr. Ojha, H. Kramsapi, P. P. Baruah, and M. Kr. Dutta, “Road network analysis of Guwahati city using GIS,” *SN Appl. Sci.*, vol. 1, no. 8, p. 906, Aug. 2019, doi: 10.1007/s42452-019-0907-4.

[12] W. Peng, G. Dong, K. Yang, and J. Su, “A Random Road Network Model and Its Effects on Topological Characteristics of Mobile Delay-Tolerant Networks,” *IEEE Trans. on Mobile Comput.*, vol. 13, no. 12, pp. 2706–2718, Dec. 2014, doi: 10.1109/TMC.2013.66.

[13] J.-P. Rodrigue, “Types of Transportation Networks and Vulnerabilities,” in *The Geography of Transport Systems*, 6th ed., New York: Routledge, 2024.

[14] D. Rivera-Royero, G. Galindo, M. Jaller, and J. Betancourt Reyes, “Road network performance: A review on relevant concepts,” *Computers & Industrial Engineering*, vol. 165, p. 107927, Mar. 2022, doi: 10.1016/j.cie.2021.107927.

[15] T. Verma, N. A. M. Araújo, and H. J. Herrmann, “Revealing the structure of the world airline network,” *Sci Rep*, vol. 4, no. 1, p. 5638, Jul. 2014, doi: 10.1038/srep05638.

[16] M. Bruneau *et al.*, “A Framework to Quantitatively Assess and Enhance the Seismic Resilience of Communities,” *Earthquake Spectra*, vol. 19, no. 4, pp. 733–752, Nov. 2003, doi: 10.1193/1.1623497.

[17] M. Omer, A. Mostashari, and R. Nilchiani, “Assessing resilience in a regional road-based transportation network,” *IJISE*, vol. 13, no. 4, p. 389, 2013, doi: 10.1504/IJISE.2013.052605.

[18] A. A. Ganin, M. Kitsak, D. Marchese, J. M. Keisler, T. Seager, and I. Linkov, “Resilience and efficiency in transportation networks,” *Sci. Adv.*, vol. 3, no. 12, p. e1701079, Dec. 2017, doi: 10.1126/sciadv.1701079.

[19] J. Zimmerman, S. Chung, G. Savant, G. L. Brown, and B. Boyd, “Quantification of coastal transportation network resilience to compounding threats from flooding and anthropogenic disturbances: A New York City case study,” *Shore & Beach*, pp. 38–44, Jun. 2023, doi: 10.34237/10091225.

[20] H. M. I. Kays, A. M. Sadri, K. K. “Muralee” Muraleetharan, P. S. Harvey, and G. A. Miller, “Exploring the Interdependencies Between Transportation and Stormwater Networks: The Case of Norman, Oklahoma,” *Transportation Research Record: Journal of the Transportation Research Board*, vol. 2678,

no. 5, pp. 491–513, May 2024, doi: 10.1177/03611981231189747.

[21] M. A. Ahmed, A. M. Sadri, A. Mehrabi, and A. Azizinamini, “Identifying Topological Credentials of Physical Infrastructure Components to Enhance Transportation Network Resilience: Case of Florida Bridges,” *J. Transp. Eng., Part A: Systems*, vol. 148, no. 9, p. 04022055, Sep. 2022, doi: 10.1061/JTEPBS.0000712.

[22] M. A. Ahmed, H. M. I. Kays, and A. M. Sadri, “Centrality-based lane interventions in road networks for improved level of service: the case of downtown Boise, Idaho,” *Appl Netw Sci*, vol. 8, no. 1, p. 2, Jan. 2023, doi: 10.1007/s41109-023-00532-z.

[23] S. Fu-li, L. Yong-lin, and Z. Yi-fan, “A Military Communication Supernetwork Structure Model for Netcentric Environment,” in *2010 International Conference on Computational and Information Sciences*, Chengdu, China: IEEE, Dec. 2010, pp. 33–36. doi: 10.1109/ICCIS.2010.16.

[24] Department of the Army, “Appendix C - Contested Deployments,” in *Army Field Manual 3-0*, 2022nd ed., Headquarters Department of the Army, 2022.

[25] SDDCTEA, “Highways for National Defense (HND),” The Official Homepage of SDDCTEA. Accessed: May 15, 2024. [Online]. Available: https://www.sddc.army.mil/sites/TEA/Functions/SpecialAssistant/Pages/HighwaysNationalDefense.aspx

[26] J. W. Cowin, “Highways for National Defense,” SDDCTEA, Aug. 2013.

[27] USDOT BTS, “Military Bases,” National Transportation Atlas Database. Accessed: Mar. 17, 2024. [Online]. Available: https://data-usdot.opendata.arcgis.com/datasets/usdot::military-bases/explore

[28] USDOT BTS, “Commercial Strategic Seaports.” Accessed: Mar. 17, 2024. [Online]. Available: https://data-usdot.opendata.arcgis.com/datasets/usdot::commercial-strategic-seaports/explore?location=35.043450,-71.937310,5.01

[29] GAO, “Defense Logistics: The Department of Defense’s Report on Strategic Seaports Addressed All Congressionally Directed Elements,” United States Government Accountability Office, GAO-13-511R, May 2013. [Online]. Available: https://www.gao.gov/assets/gao-13-511r.pdf

[30] J. L. Gross and J. Yellen, *Graph theory and its applications*, 2nd ed. in Discrete mathematics and its applications. Boca Raton: Chapman & Hall/CRC, 2006.

[31] L. Lu and M. Zhang, “Edge Betweenness Centrality,” in *Encyclopedia of Systems Biology*, W. Dubitzky, O. Wolkenhauer, K.-H. Cho, and H. Yokota, Eds., New York, NY: Springer New York, 2013, pp. 647–648. doi: 10.1007/978-1-4419-9863-7_874.

[32] U. Brandes, “On variants of shortest-path betweenness centrality and their generic computation,” *Social Networks*, vol. 30, no. 2, pp. 136–145, May 2008, doi: 10.1016/j.socnet.2007.11.001.

[33] L. C. Freeman, “A Set of Measures of Centrality Based on Betweenness,” *Sociometry*, vol. 40, no. 1, p. 35, Mar. 1977, doi: 10.2307/3033543.